\documentclass[aps,prd,twocolumn,superscriptaddress,nofootinbib]{revtex4-2}

\usepackage[T1]{fontenc}
\usepackage[utf8]{inputenc}
\usepackage[american]{babel}
\usepackage{epsfig}
\usepackage{xspace}
\usepackage{graphicx}
\usepackage{booktabs}
\usepackage{multirow}
\usepackage{dcolumn}
\usepackage{amsmath}
\usepackage{subfigure}
\usepackage{mathtools}
\usepackage{amsfonts}
\usepackage{amssymb}
\usepackage{epstopdf}
\usepackage{siunitx}
\usepackage{braket}
\usepackage{enumitem}
\usepackage{soul}
\usepackage[table]{xcolor}
\usepackage{color}
\usepackage{pifont}
\usepackage[normalem]{ulem}

\usepackage{hyperref}
\hypersetup{
    colorlinks=true, 
    linkcolor=blue, 
    citecolor=magenta}

\begin{document}

\title{Amplifying muon-to-positron conversion in nuclei with ultralight dark matter}

\author{Purushottam Sahu}
\email{purushottam.sahu@iitb.ac.in}
\affiliation{Department of Physics, Indian Institute of Technology Bombay, Mumbai 400076, India}

\author{Manibrata Sen}
\email{manibrata@iitb.ac.in}
\affiliation{Department of Physics, Indian Institute of Technology Bombay, Mumbai 400076, India}


\begin{abstract}
We present an analysis of the lepton-number and lepton-flavour-violating process of muon-to-positron conversion $\mu^- + N \rightarrow e^+ + N'$, in the presence of an ultralight scalar dark matter (ULSDM) field which couples to neutrinos. The ULSDM contributes to the effective off-diagonal Majorana mass $ m_{\mu e}$, therefore amplifying the rate of muon-to-positron conversion to experimentally observable levels. Using existing bounds from SINDRUM II, COMET, and Mu2e experiments, we derive novel constraints on the flavour-off-diagonal couplings of neutrinos to ULSDM. Our work reveals that upcoming experiments can provide stronger sensitivity to these new couplings than bounds arising from cosmological surveys and terrestrial experiments. 
\end{abstract}

\maketitle

\section{Introduction}
\label{sec1}
The Standard Model (SM) of particle physics presents a robust framework for the electroweak and strong sectors and has been successfully validated across different experiments. Despite its glorious success, there are some glaring loopholes: the SM fails to account for several key phenomena, such as the origin of the neutrino mass and the possibility of lepton number violation (LNV), the presence of dark matter (DM), and the matter-antimatter asymmetry of the Universe, among others.
The discovery of neutrino oscillations implies that neutrinos have masses and they mix~\cite{Cleveland:1998nv, Kamiokande-II:1989hkh, SAGE:1999nng, GALLEX:1998kcz, Super-Kamiokande:1998qwk, SNO:2001kpb, Gajewski:1992iq, Kamiokande-II:1992hns,  Kamiokande:1994sgx, Super-Kamiokande:1998kpq}, but the underlying mechanism for both remains unclear. It is widely believed that at the heart of this puzzle lies the mechanism of LNV. Majorana neutrinos, violating the total LN by two units, could enable processes like neutrinoless double beta decay $(0\nu\beta\beta)$~\cite{Schechter:1981bd,Nieves:1984sn,Bilenky:2001rz,Patra:2023ltl}  and $\mu^- \to e^+$ conversions~\cite{Simkovic:2000ma,Divari:2002sq,Mu2e:2014fns,Kuno:2015tya,Geib:2016atx,COMET:2018auw,Lee:2021hnx}. 
As a result, several theoretical and experimental efforts are underway to understand these processes and probe the nature of LNV.
For $0\nu\beta\beta$, various isotopes such as $^{76}$Ge, $^{136}$Xe, and $^{130}$Te have been used in experiments to push the half-life limits to around $10^{26}$ years~\cite{KamLAND-Zen:2016pfg,KamLAND-Zen:2022tow,GERDA:2020xhi,EXO-200:2019rkq,CUORE:2021mvw,PhysRevLett.123.032501,Majorana:2019nbd,Majorana:2022udl}. The non-observation of $0\nu\beta\beta$ thus far places significant constraints on LNV and models that incorporate Majorana neutrino masses.

The other rare process of LNV, muon-to-positron conversion in a nucleus, $ \mu^- + (A,Z) \rightarrow e^+ + (A,Z-2)  $  is also a total lepton flavour violating (LFV) transition with $\Delta L_\mu = -1$, and $\Delta L_e = +1$. In the SM with massless neutrinos, such processes are strictly forbidden. Even when Majorana neutrino masses are incorporated, the expected branching ratios are suppressed to the level of $\sim 10^{-40}$ due to helicity suppression and the smallness of neutrino masses \cite{Simkovic:2000ma,Domin:2004tk}. 

However, in many beyond-the-Standard-Model (BSM) scenarios, such as those involving heavy RH neutrinos or new mediator particles, this rate can be significantly enhanced~\cite{Simkovic:2001fs,Domin:2004tk,Berryman:2016slh,Geib:2016daa,Sato:2022vny}. Furthermore, model-independent effective field theory approach has also been applied to study the BSM implications of this process~\cite{Babu:2001ex,deGouvea:2007xp,Angel:2012ug,deGouvea:2019xzm, Haxton:2024lyc}. Any observable signal in this particular channel can be a clear indication of the presence of new physics. In fact, in certain cases, as we will see, $\mu^- \to e^+$ presents a unique probe of the flavour structure of new physics as well. This will be complementary to probes from other charged lepton-flavour-violating (CLFV) processes like $\mu \to e\gamma$ and $\mu \to 3e$, which have been extensively probed by MEG \cite{MEG:2016leq} and SINDRUM \cite{Kaulard:1998rb}, respectively.

Experimentally, the SINDRUM II experiment places the strongest bound on $\mu^- \to e^+$, with a limit of $\text{Br}(\mu^- + \text{Ti} \to e^+ + \text{Ca}) < 1.7 \times 10^{-12}$ at 90\% confidence level~\cite{Kaulard:1998rb}. Next-generation experiments such as COMET~\cite{COMET:2018auw} and Mu2e~\cite{Mu2e:2014fns} will improve this sensitivity by several orders of magnitude, thereby probing branching ratios as low as $10^{-17}$. Nevertheless, these processes are plagued by different backgrounds, the most important of which is the radiative muon capture (RMC), especially near the photon energy endpoint~\cite{Lee:2021hnx}. Different target materials, such as titanium and calcium, are considered to mitigate this issue.

It is possible that through a SM portal, DM can play a direct role in influencing such processes. One particularly interesting scenario involves an ultralight scalar dark matter (ULSDM), with masses in the range $10^{-22} - \mathcal{O}(1)$ eV~\cite{Hu:2000ke,Ferreira:2020fam}. Under certain circumstances, for low enough masses, ULSDM can be treated as a coherent classical field, with quantum fluctuations suppressed by inverse powers of the large occupation number of these particles. The fascinating consequences of ULSDM coupling to neutrinos have been explored extensively in the literature, focussing on their signatures on neutrino oscillation experiments~\cite{Berlin:2016woy,Brdar:2017kbt,Capozzi:2018bps,Dev:2020kgz,Losada:2021bxx}, beta decay~\cite{Huang:2022wmz}, neutrino mass~\cite{Davoudiasl:2018hjw,Chun:2021ief, Dev:2022bae, Sen:2023uga, Goertz:2024gzw} as well as cosmology and astrophysics~\cite{Hui:2016ltb, Choi:2019zxy,Choi:2020ydp,Sen:2024pgb, Martinez-Mirave:2024dmw}. 
Such a neutrinophilic ULSDM could stimulate and enhance the rates of such LN-violating processes, thereby making them potentially detectable in current and future experiments. Previous works along these lines have explored such a mechanism to enhance the rate of $0\nu\beta\beta$ decay~\cite{Huang:2021kam}. In fact, it was shown that if the ULSDM carries LN, it can also lead to $0\nu\beta\beta$ decay even in the absence of LNV~\cite{Graf:2023dzf}. 

While $0\nu\beta\beta$ decay probes flavour diagonal coupling of ULSDM to neutrinos, our work, for the first time, probes flavour off-diagonal couplings through the transitions $\mu^- \to e^+$. We investigate how the presence of a ULSDM field $\phi$, coupled to neutrinos via an effective interaction term $g_{\mu e} \nu_e \nu_\mu\phi^*$, can modify the effective Majorana masses appearing in $\mu^- \to e^+$ conversion. This coupling induces an additional contribution to the process, thereby enhancing the transition rate. We compute the expected branching ratios and compare them to experimental bounds from SINDRUM II \cite{SINDRUMII:1998mwd}, and projected sensitivities from COMET \cite{COMET:2018auw}, and Mu2e \cite{Mu2e:2014fns}. We show that future experiments can probe large portions of the $g_{\mu e} - m_\phi$ parameter space, even beyond cosmological constraints. We argue that even in the foreseeable future, $\mu^- \to e^+$ experiments can provide one of the strongest direct constraints on such flavour-off-diagonal couplings of neutrinos with ULSDM, competetive with indirect constraints from astrophysical and cosmological sources.
In this work, we focus on the off-diagonal coupling $g_{\mu e}$, which allows one to isolate in a model-independent way the sensitivity of $\mu^-\to e^+$ conversion to flavour off-diagonal neutrinophilic ULSDM interactions. Flavour-diagonal couplings, in particular in the $ee$ sector, are strongly constrained by neutrinoless double-beta decay and provide a complementary probe of the diagonal entries of the same coupling matrix.

Our work is structured as follows. In Sec.\,\ref{sec2}, we discuss the possibility of the vanilla nuclear muon-to-positron conversion in the absence of a ULSDM. Sec.\,\ref{sec3} discusses how this scenario changes when a neutrinophilic ULSDM is added to the picture. The presence of existing constraints on such a coupling is discussed in Sec.\,\ref{sec4}, before concluding in Sec.\,\ref{sec5}.
\section{Nuclear Muon-to-Positron Conversion}
\label{sec2}
\begin{figure}[!t]
\centering
        \includegraphics[width=0.4\textwidth]{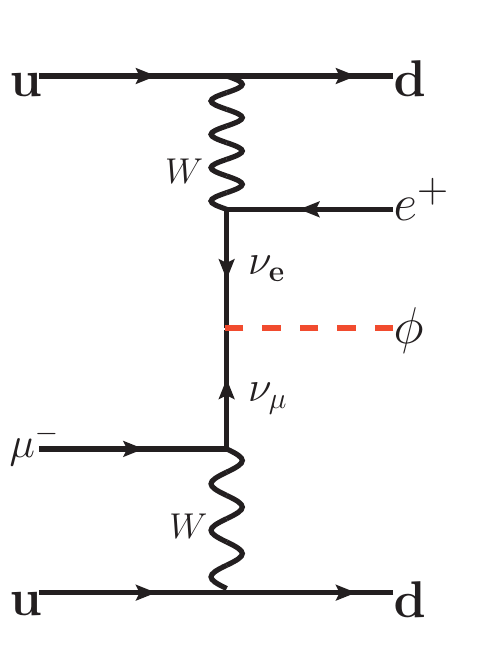}
    \caption{Feynman diagram for $\mu^- \rightarrow e^+$ conversion in nuclei in the presence of LN carrying ULSDM $\phi$ (dashed red line). In the absence of the ULSDM, the interaction is LNV and proceeds only with a mass insertion between the neutrino flavours (no $\phi$ interaction) .}
    \label{fig:FD-mue}
\end{figure}

 Nuclear muon-to-positron conversion proceeds via the capture of a muon in an atom, thereby forming a muonic atom, followed by an incoherent transition into a positron: $\mu^- + (A,Z) \rightarrow e^+ + (A,Z-2)$.  Analogous to $0\nu\beta\beta$, this process can proceed via the exchange of a virtual Majorana neutrino, similar to the process shown in Fig.\,\ref{fig:FD-mue} (regular mass insertion between neutrino flavours, without the interaction involving $\phi$). Within the framework of the SM extended to include nonzero Majorana neutrino masses, the amplitude is suppressed by the smallness of the neutrino masses and the chiral structure of weak interactions. The branching ratio for the process is given by \cite{Simkovic:2000ma,Domin:2004tk}:
\begin{equation}
    R_{\mu^- \to e^+} = \frac{\Gamma(\mu^- \to e^+)}{\Gamma(\mu^- \to \nu_\mu)} \approx 2.6 \times 10^{-22}\,\frac{|m_{\mu e}|^2}{m_e^2}|M_{\text{n}}|^2,
    \label{Eq:mutoerate}
\end{equation}
where $m_{\mu e}$ is the effective Majorana mass controlling the process, $m_e$ is the electron mass, $|M_{\text{n}}|$ is the nuclear matrix element (NME). The value of \(|M_{\text{n}}|\) varies with the nucleus: for titanium, estimates range from 0.03 to 0.5~\citep{Domin:2004tk,Berryman:2016slh}. For the aluminium-to-sodium transition targeted by COMET and Mu2e, a benchmark value of 0.1 is commonly used, though model-dependent uncertainties remain~\cite{Berryman:2016slh}.

The effective Majorana mass arises from the underlying LNV operator and is defined as
\begin{equation}
m_{\mu e}  = \sum_{i=1}^3 U_{ei} U_{\mu i} m_i e^{i \alpha_i},
\label{eq:mass_vacuum}
\end{equation}
where \(U_{\alpha i}\) are elements of the Pontecorvo-Maki-Nakagawa-Sakata (PMNS) matrix, \(m_i\) are the neutrino mass eigenvalues, and \(\alpha_i\) are the Majorana phases. This mass is sensitive to the neutrino mass ordering, mixing angles, and CP-violating phases.  Importantly, the structure of $ m_{\mu e}$ differs from that of $m_{ee}$ in $0\nu\beta\beta$, due to its dependence on off-diagonal combinations of PMNS elements. This introduces a more pronounced dependence on the Dirac and Majorana CP-violating phases, as well as the neutrino mass ordering. For example, in the inverted mass-ordering (IO), destructive interference among terms can suppress $ m_{\mu e}$ below the levels found in the normal mass ordering (NO).

\begin{figure}[!t]
\centering
        \includegraphics[width=0.5\textwidth]{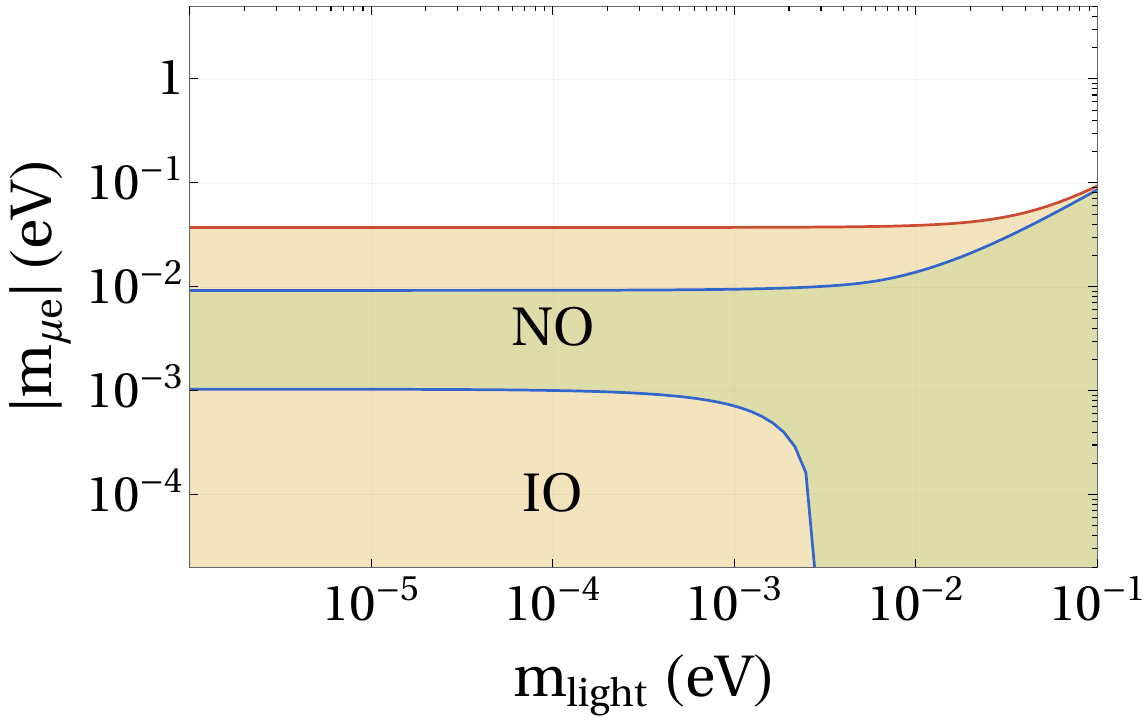}
    \caption{$3\sigma$ allowed regions for the effective Majorana mass $| m_{\mu e}|$ (in eV) versus the lightest neutrino mass $m_{\text{light}}$ (in eV), for Normal Ordering (NO, green band, labeled “NO”) and Inverted Ordering (IO, golden band, labeled “IO”), computed using 3$\sigma$ oscillation parameters. The constraints arising from experiments are summarised in Table~\ref{tab:exp_bounds}.}
    \label{fig:majorana-mue}
\end{figure}

Fig.\,\ref{fig:majorana-mue} shows the $3\sigma$ allowed range of $|m_{\mu e}|$ as a function of the lightest neutrino mass $m_{\text{light}}$ for both NO (green band) and IO (golden band).
All neutrino oscillation parameters are varied within their $3\sigma$ ranges~\cite{deSalas:2020pgw}, while the Majorana phases are scanned over the full range (0,2$\pi$)~\cite{deSalas:2020pgw}.
The nature of the allowed region can be broadly explained through the following regimes of 
$m_{\text{light}}$.

\textbf{Hierarchical regime} ($m_{\text{light}} \lesssim 10^{-3}$~eV):  
For NO, where $m_1 \approx 0$, the effective mass lies in the range $|m_{\mu e}| \sim (3$--$8) \times 10^{-3}$~eV. In IO, where $m_3 \approx 0$, nearly degenerate $m_1 \approx m_2$ allows for significant cancellation, yielding $| m_{\mu e}| \sim 10^{-5}$~eV.

\textbf{Transition regime} ($m_{\text{light}} \sim 0.01$--$0.1$~eV):  
As mass splittings become subdominant, interference between terms can suppress $| m_{\mu e}|$ further, particularly for NO with tuned CP phases. IO remains suppressed throughout due to its nearly degenerate mass pattern.

\textbf{Degenerate regime} ($m_{\text{light}} \gtrsim 0.1$~eV): Both orderings yield $|m_{\mu e}| \propto m_{\text{light}}$, with possible cancellations near $10^{-5}$ eV.


With this, one needs access to the nuclear matrix element values to estimate the rate of the process.  
Considering \(|M_{\text{n}}| = 0.1\)~\cite{Domin:2004tk,Berryman:2016slh},  Eq.\,\ref{Eq:mutoerate} becomes
 
\begin{equation}
R_{\mu^-\to e^+} \simeq 10^{-35} \left( \frac{|m_{\mu e}|}{\text{eV}} \right)^2\,,
\label{eq:rate_final}
\end{equation}
where $|m_{\mu e}|$ is measured in eV. This highlights the quadratic dependence of the rate on the effective off-diagonal Majorana mass, $m_{\mu e}$, emphasising its central role in determining experimental sensitivity.
Experimental limits on $R_{\mu^- \to e^+}$ can be translated to limits on $m_{\mu e}$. The limits obtained using current and projected sensitivities are summarised in Table \ref{tab:exp_bounds}.

\begin{table}[!h]
\centering
\begin{tabular}{|l|c|c|}
\hline
Experiment & $R_{\mu e}$ & $| m_{\mu e}|$ (eV) \\
\hline
 &  & \\
SINDRUM II~\cite{SINDRUMII:1998mwd} & $< 1.7 \times 10^{-12}$ & $< 4 \times 10^{11}$ \\
COMET Phase-I~\cite{COMET:2018auw} & $< 10^{-14}$ & $< 3 \times 10^{10}$ \\
Mu2e~\cite{Mu2e:2014fns} & $< 10^{-16}$ & $< 3 \times 10^{9}$ \\
COMET Phase-II~\cite{COMET:2018auw} & $< 6 \times 10^{-17}$ & $< 2 \times 10^{9}$ \\
 &  & \\
\hline
\end{tabular}
\caption{Experimental bounds on the conversion rate $R_{\mu e}$ and the effective Majorana mass $|m_{\mu e}|$ for $\mu^- \to e^+$.}
\label{tab:exp_bounds}
\end{table}

Even in the most optimistic case, the experimental sensitivities extend down to $\sim 10^9$~eV only. This enormous gap underscores the limitations of light Majorana neutrino exchange in producing observable $\mu^- \to e^+$ rates. This motivates the search for BSM mechanisms. In this work, we explore how ultralight scalar dark matter can dynamically enhance $| m_{\mu e}|$ through coherent background effects, potentially bringing it within reach of Mu2e and COMET Phase-II.

\section{Effects of Ultralight Scalar Dark Matter}
\label{sec3}
ULSDM has been proposed as an alternative to cold or warm dark matter to account for certain discrepancies that exist between small-scale structure observations and predictions from numerical N-body cosmological simulations~\cite{Hu:2000ke}. Due to low masses, such fields can be treated as a classical background field, with quantum fluctuations suppressed by the large occupation number of the fields. Such ultralight fields can account for part or the whole of the DM relic density of the Universe.

\begin{figure}[t!]
    \centering
    \includegraphics[width=1.0\linewidth]{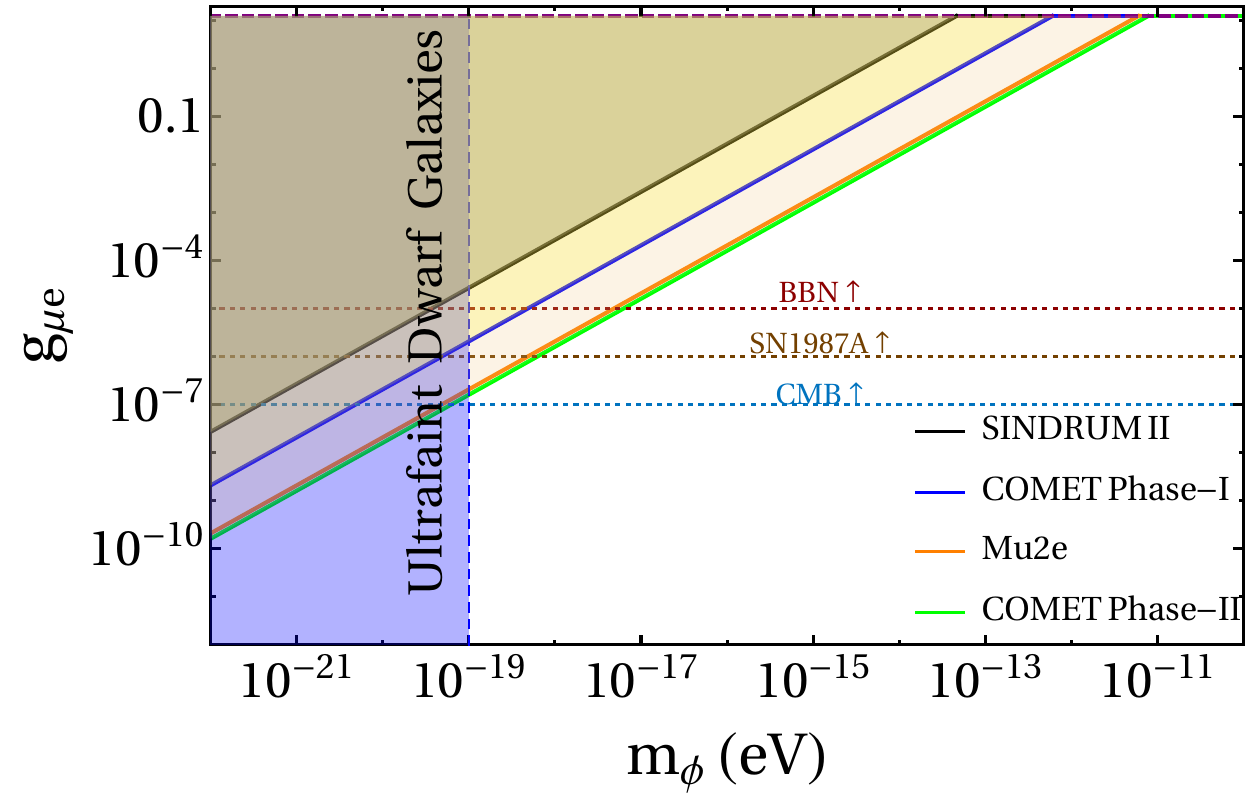}
    \caption{Exclusion regions in the \(g_{\mu e}\)--\(m_\phi\) parameter space. Shaded regions are excluded by SINDRUM II (grey), COMET Phase-I (cyan), Mu2e (orange), and COMET Phase-II (green). Dashed line shows perturbative limit \(g_{\mu e} = 4\pi\). The shaded blue region shows constraints from ultrafaint dwarf galaxy heating~\cite{Dalal:2022rmp}. Horizontal dotted lines labeled “BBN$\uparrow$”, “SN1987A$\uparrow$”, and “CMB$\uparrow$” indicate 
approximate cosmological and astrophysical upper bounds on $g_{\mu e}$ from Big Bang Nucleosynthesis, 
core–collapse supernova cooling, and CMB neutrino free streaming, respectively; the arrows denote 
that the region above each line is excluded at the order of magnitude level (see detailed discussions in Sec.\,\ref{sec4}). These cosmological and 
astrophysical bounds are indirect and somewhat model dependent, while the $\mu^- \to e^+$ limits from 
SINDRUM II, COMET, and Mu2e provide direct laboratory constraints on $g_{\mu e}$.}
    \label{fig:constraints}
\end{figure}

In this study, we consider an interaction between ULSDM and neutrinos through an effective LFV dimension-six operator,
\begin{equation}
    \mathcal{L}_{\alpha\beta}\supset \frac{y_{\alpha\beta}}{\Lambda^2} (L_{\alpha}H) (L_{\beta} H) \phi^* + \text{h.c.},
\label{eq:dim6}    
\end{equation}
where $\phi^*$ is the ULSDM field carrying LN of two units, \textcolor{blue}{$y_{\alpha,\beta}$ }is a dimensionless coupling constant, $L_{\alpha,\beta}$ are the SM lepton doublets. This operator violates LF but preserves total LN. More generally, this operator can be viewed as an effective term arising from integrating out heavier degrees of freedom at some scale $\Lambda$. For this particular study, we focus on $\alpha=\mu,\,\beta=e$. After electroweak symmetry breaking, the Lagrangian becomes 
\begin{equation}
    \mathcal{L}_{\mu e}\supset g_{\mu e} \nu_\mu \nu_e \phi^* + \text{h.c.}\,,
    \label{eq:Lagrangian}
\end{equation}
where $g_{\mu e}= y_{\mu e} v^2/\Lambda^2$. Generic ultraviolet completions of this operator have been considered in~\cite{Berryman:2016slh}.
Numerically, one finds $\Lambda \simeq v \sqrt{y_{\mu e} / g_{\mu e}}$, so that the couplings $g_{\mu e} \sim 10^{-5}-10^{-10}$ shown in Fig.~\ref{fig:constraints} correspond to a new-physics scale $\Lambda \sim (10^5 - 10^7) \text{ GeV}$ for $y_{\mu e} = \mathcal{O}(1)$. Such scales are natural in UV completions where the operator in Eq.~(\ref{eq:dim6}) is generated by integrating out heavy mediators.

In the coherent field limit where the ULSDM behaves as a classical scalar field, it can be expressed as $\phi(t) = \phi_0 \cos(m_\phi t)$, with $\phi_0 = \sqrt{2 \rho_\phi}/m_\phi$. Here $\rho_\phi = 0.3$ GeV/cm$^3$ is the local DM density of the Universe, and $m_\phi$ denotes the mass of the scalar field. A simple estimation shows that the time modulation associated with the scalar field has a time period, $\tau_\phi \simeq 1\,{\rm yr}\,\left(10^{-22}\,{\rm eV}/m_\phi\right)$. This leads to a time-varying contribution to the effective Majorana mass for $\mu^- \to e^+$ conversion:
\begin{equation}
     m_{\mu e}(t) =  (m_{\mu e})_{\text{vac}} + g_{\mu e} \phi_0 \cos(m_\phi t)\,,
\end{equation}
where $(m_{\mu e})_{\text{vac}}$ is the vacuum contribution defined by Eq.\,\ref{eq:mass_vacuum}.

If the time modulation is way faster than the experimental time scales, the effect is averaged out. This leads to an enhancement of the effective off-diagonal Majorana mass-squared, given by 
\begin{equation}
    \langle| m_{\mu e}|^2 \rangle = \langle |(m_{\mu e})_{\text{vac}}|^2 \rangle + \frac{1}{2}(g_{\mu e} \phi_0)^2.
    \label{eq:enhancement}
\end{equation}
This acts like an additional contribution to the $\mu^- \to e^+$ process, stimulated by the presence of the uniform ULSDM. As a result, the ULSDM field acts as a cosmic amplifier, thereby potentially increasing the signal to experimentally accessible regions. 
In addition to the coherent-background effect considered here, one could in principle also have real ULSDM emission in the process $\mu^- N \to e^+ N' \phi$ by attaching an external $\phi$ leg to the internal neutrino line. Such a three-body process is, however, parametrically suppressed by an additional power of the coupling, with $\Gamma_{\mu^- N \to e^+ N' \phi} \propto g_{\mu e}^4$, and by phase space, compared to the coherent enhancement $\Gamma_{\mu^- N \to e^+ N'} \propto g_{\mu e}^2$. For the small couplings of interest in Fig.~\ref{fig:constraints}, the rate for real $\phi$ emission is therefore completely negligible.
Furthermore, since $\phi$ carries LN, this additional contribution due to ULSDM can happen even if there is no LNV, i.e., even when $(m_{\mu e})_{\text{vac}}$ is zero. This is depicted in Fig.\,\ref{fig:FD-mue} where the $\phi$ (red line) can stimulate these processes even in the absence of LNV.
Therefore, this weakens the claim that the observation of $\mu^- \to e^+$ will give rise to a smoking gun signature of LNV. On the other hand, non-observation of such a process enables us to translate experimental bounds on $R$ into exclusion regions in the $g_{\mu e}$--$m_\phi$ plane.

Fig.\,\ref{fig:constraints} displays the exclusion contours in the \( g_{\mu e} \)--\( m_\phi \) parameter space for the ULSDM model, covering scalar masses from \( 10^{-22} \) to \( 10^{-10} \, \text{eV} \). Solid curves represent the constraints from SINDRUM II, COMET Phase-I, Mu2e, and COMET Phase-II, derived from their respective upper bounds on the \( \mu^- \to e^+ \) branching ratio. 
The blue-shaded region below \( m_\phi = 10^{-19} \, \text{eV} \) denotes the bounds coming from heating of ultrafaint dwarf galaxies~\cite{Dalal:2022rmp}. The upper limit is set by perturbativity constraints  $ g_{\mu e} < 4\pi$. These complementary probes together carve out a window of viable parameter space uniquely accessible to LFV nuclear processes. 
Each exclusion curve in Fig.~\ref{fig:constraints} corresponds to the set of $(g_{\mu e}, m_\phi)$ values for which the ULSDM-induced effective mass $m_{\mu e}$ yields a conversion rate that saturates the corresponding experimental sensitivity. Points above a given curve are excluded, while points below predict a smaller branching ratio. In the region where the ULSDM contribution dominates, $\langle |m_{\mu e}|^2 \rangle \simeq (g_{\mu e} \phi_0)^2 / 2$, so that Eq.~(\ref{eq:rate_final}) implies that parameter points just below the Mu2e or COMET Phase-II curves give $R_{\mu^-\to e^+} \sim 10^{-17}-10^{-16}$, i.e., within an order of magnitude of the projected sensitivities.

Throughout our analysis, we assume that the time-dependent modulation induced by the coherent ULSDM field can be treated by averaging over oscillation cycles when the scalar mass satisfies $m_\phi \gg 1/\tau_{\rm exp}$, where $\tau_{\rm exp}$ denotes the effective experimental integration time. However, in the intermediate regime where the ULSDM oscillation period becomes comparable to or longer than the integration time (e.g., $m_\phi\sim 10^{-22}\,\mathrm{eV}$), this simple time-averaging prescription may no longer hold. In such cases, the conversion rate could exhibit temporal modulations over months to years, potentially requiring a dedicated analysis to model the effective sensitivity and statistical power of the experiments. Although currently undetectable due to limited timing precision, future advancements may allow correlation of the conversion signal with the dark matter oscillation frequency, providing a smoking-gun signature of the ULSDM origin.

The ULSDM background also induces a tiny invisible decay channel $h \to \nu_e \nu_\mu$ via the effective coupling $(g_{\mu e} \phi_0 / v)\, h \nu_e \nu_\mu$. Using $\phi_0 \simeq \sqrt{2\rho_\phi}/m_\phi$ with $\rho_\phi \simeq 0.3~\text{GeV}/\text{cm}^3$ and imposing the oscillation bound $g_{\mu e}\phi_0 \lesssim 0.008~\text{eV}$, one finds $(g_{\mu e}\phi_0 / v) \lesssim 10^{-14}$, which implies an invisible branching ratio $\text{Br}(h \to \nu_e \nu_\mu) \ll 10^{-20}$. This is many orders of magnitude below the current collider limit on invisible Higgs decays (of order $10\%$), and can be safely neglected.

If diagonal entries of the ULSDM--neutrino coupling matrix are present in addition to the off-diagonal term, they would generically induce contributions to neutrinoless double-beta decay ($0\nu\beta\beta$), both in the presence and in the absence of explicit lepton-number violation~\cite{Huang:2021kam,Graf:2023dzf}. The analysis presented here should therefore be viewed as complementary to those works: $0\nu\beta\beta$ predominantly constrains flavour-diagonal couplings, while $\mu^-\to e^+$ conversion is directly sensitive to the flavour off-diagonal entry.

\section{Other constraints}
\label{sec4}
The couplings considered here significantly influence early-universe cosmology through interactions with the neutrino sector. The primary constraints arise from excess radiation during the Big Bang Nucleosynthesis (BBN) epoch and modifications of the Cosmic Microwave Background (CMB). Additionally,  terrestrial and astrophysical environments offer complementary probes. We discuss these constraints in this section.
\subsection{Thermalization and $N_\text{eff}$}
The coupling $\phi^* \, \nu_\mu \nu_e$ can lead to the thermal production of $\phi$ around the BBN epoch, increasing the number of relativistic degrees of freedom,$N_\text{eff}$~\cite{Wagoner1967,Boesgaard1985}. For very low masses of $\phi$, the inverse decay process $\nu_e + \nu_\mu \to \phi$ is inefficient and the $2\to 2$ production via $\nu_e + \bar{\nu}_\mu \to \phi + \phi^*$ becomes relevant~\cite{Huang:2017egl}. Requiring that  $\Gamma \sim g_{\mu e}^4 T < H \sim T^2 / M_\text{Pl}$ at $T \sim 10$ MeV, leads to $g_{\mu e} \lesssim 10^{-5}\,$
in order to prevent $\phi$ from reaching equilibrium and contributing significantly to $N_\text{eff}$. 
\subsection{Neutrino Free-Streaming and the CMB}
The observation of the acoustic peaks in the CMB power spectrum implies that neutrinos must free-stream by the epoch of recombination, i.e., $T \sim 1$ eV. However, additional neutrino scattering channels, such as $\nu_\mu + \nu_e \leftrightarrow \nu_\mu + \nu_e$, or the inverse decay process $\nu_\mu + \nu_e \leftrightarrow \phi$ can lead to a smoothing of the neutrino anisotropic stress~\cite{Cyr-Racine:2013jua,Oldengott:2017fhy, Kreisch:2019yzn, Das:2023npl}. Requiring that this rate of these processes falls below the Hubble parameter at $T \sim 1$ eV gives the bound:
$ g_{\mu e} \lesssim 10^{-7}$. Additional constraints can also arise from the contribution to the additional neutrino mass at the recombination epoch~\cite{Brdar:2017kbt,Huang:2021kam,Sen:2024pgb}.
\subsection{Core-collapse supernova Bounds}
In a core-collapse supernova (CCSN), the neutrino density is high enough to produce the ULSDM on-shell through $\bar{\nu}_\mu\to  \nu_e + \phi^*$~\cite{Farzan:2002wx}. This can lead to additional cooling channels if $\phi$ escapes freely from the SN core, and lead to a reduction of the neutrino luminosity~\cite{Farzan:2002wx,Kachelriess:2000qc,Brune:2018sab}. To avoid shortening the observed duration of the SN1987A neutrino signal, such energy loss must be suppressed. 
A detailed calculation of the cooling rate yields~\cite{Farzan:2002wx} $  g_{\mu e} \lesssim 10^{-6}$,
assuming that $\phi$ is weakly coupled enough to escape without thermalizing. If $\phi$ is trapped, this bound is relaxed, though more detailed SN simulations would be required to quantify the limit. Furthermore, such dynamic neutrino masses can also show up in the spectra of the diffuse supernova neutrino background~\cite{deGouvea:2022dtw,Perez-Gonzalez:2025qjh}.
\subsection{Neutrino Oscillation Modulation}
For very low mass scalar fields, the ULSDM may maintain its coherence over cosmological timescales. This can lead to a time-dependent correction to the neutrino mass matrix. Such off-diagonal mass terms can modify neutrino oscillation probabilities or induce apparent spectral distortions~\cite{Berlin:2016woy, Dev:2020kgz, Dev:2022bae}. Demanding that the effective contribution to the neutrino mass is smaller than the solar mass-squared difference yields $g_{\mu e} \, \phi_0 \lesssim \sqrt{\Delta m^2_{\rm sol}} \sim 0.008\,{\rm eV}\,$. However, note that this limit only applies in the very low $m_\phi$ regime, $m_\phi\lesssim (10^{-18} - 10^{-19}) \,{\rm eV}$. Typically, for larger values of $m_\phi$, the oscillation bounds weaken due to averaging out of the time modulations. On the other hand, $\mu^- \to e^+$ conversion is sensitive to the time-averaged quadratic coupling induced by an oscillating ULDM field, thereby probing a complementary and unconstrained region of parameter space. Alternatively, this bound can be evaded by introducing Majorana mass terms for the $\nu_e$ and $\nu_\mu$, along with additional interactions of the type in Eq.\,(\ref{eq:Lagrangian}) in the electron and muon sectors. This leads to a see-saw-like structure, where large diagonal terms can prevent off-diagonal couplings from generating large mass shifts to the active neutrinos.
In summary, for $m_\phi \lesssim (10^{-18} - 10^{-19})  \text{ eV}$, the conservative requirement $g_{\mu e} \phi_0 \lesssim \sqrt{\Delta m_{\text{sol}}^2}$ guarantees that the time-dependent off-diagonal mass term does not significantly modify established neutrino oscillation probabilities. For larger $m_\phi$, the rapid oscillations average out and the effect of ULSDM on macroscopic oscillations has to be treated on a different footing (check~\cite{Dev:2022bae} for details). The parameter region probed by $\mu^-\to e^+$ conversion and shown in Fig.~\ref{fig:constraints} is always taken to be consistent with these oscillation constraints.

\subsection{Charged Lepton Flavour Violation}
The coupling $\phi^* \, \nu_\mu \nu_e$ breaks LF but conserves total LN. Since processes like $\mu \to e \gamma$ conserve LN, this flavour-violating coupling could potentially contribute at higher orders only. However, because $\phi$ does not couple to charged leptons, the 1-loop diagrams involving $\phi$ and neutrinos do not generate such dipole operators without additional LNV or mixing with charged sectors.
As such, processes like $\mu \to e \gamma$, $\mu \to 3e$, and $\mu$--$e$ conversion remain highly suppressed and do not pose additional constraints. For a detailed discussion on the effect of ultralight leptophilic scalars on CLFV processes, see~\cite{Escribano:2020wua,Bigaran:2025uzn}.

\section{Conclusion}
\label{sec5}
We have explored a new enhancement mechanism for $\mu^- \to e^+$ conversion in an atomic nucleus. This rare capture process remains a smoking gun signal for total LNV and LFV. However, even within minimal extensions of the Standard Model involving Majorana neutrino masses, the rate of this process is hopelessly small and well beyond the current experimental sensitivities. 

We consider a scenario where this process receives an amplification due to the presence of a neutrinophilic ultralight scalar dark matter (ULSDM). The specific coupling of interest in our study involves a ULSDM coupled to a $\nu_e$ and a $\nu_\mu$ in a LN conserving manner. Such a ULSDM with a LFV coupling to neutrinos can lead to an additional channel for the process, thereby providing the necessary enhancement to reach experimental sensitivities. Furthermore, due to the LN carried by the ULSDM, the presence of this coupling can lead to a $\mu^- \to e^+$ conversion, even in the absence of LNV.
Conversely, the current and projected bounds on $\mu^- \to e^+$ conversion can be used to place constraints on such an LFV coupling of neutrinos to ULSDM -- the first of its kind.

By calculating the contribution to the effective Majorana mass $ m_{\mu e}$ and incorporating a time-varying component from the scalar field, we derive new constraints on the neutrinophilic ULSDM. We find that the ULSDM contribution will dominate the conversion rate, and hence current and future experiments can probe the parameter space for light scalar masses $m_\phi \lesssim 10^{-11}$ eV and couplings as low as $g_{\mu e} \sim 10^{-10}$.
Such values are well within the reach of current and future experiments, including COMET Phase-II and Mu2e. This will allow us to place some of the strongest model-independent constraints on such LFV coupling of ULSDM with neutrinos.

Our analysis places the process in the broader context of beyond-Standard-Model searches, alongside neutrinoless double-beta decay and $\mu \to e\gamma$. These results further establish $\mu^- \to e^+$ as a competitive and complementary probe of ultralight dark matter, bridging the frontiers of neutrino physics, lepton flavour violation, and dark sector phenomenology.

\bigskip

\section*{Acknowledgement}
We would like to thank S. Umasankar for useful discussions. PS thanks the Ministry of Education, Government of India for financial support through the Institute of Eminence funding to IIT Bombay. MS acknowledges support from the Early Career Research Grant by Anusandhan National Research Foundation (project number ANRF/ECRG/2024/000522/PMS). The authors also acknowledge support of the Institut Henri Poincaré (UAR 839 CNRS-Sorbonne Université), and LabEx CARMIN (ANR-10-LABX-59-01). 
\bibliographystyle{apsrev4-2}
\bibliography{references1}

\end{document}